\documentclass[aps, prb,
preprint,amsmath,amssymb,showpacs,groupedaddress]{revtex4}
\usepackage{graphicx}
\graphicspath{{../}}
\usepackage{epstopdf}
\begin{document}

\preprint{}

\title{Flexoelectric effect in finite samples}

\author{Alexander K. Tagantsev}
\affiliation{Ceramics Laboratory, Swiss Federal Institute of Technology (EPFL), CH-1015 Lausanne, Switzerland}

\author{Alexander S. Yurkov}
\affiliation{644076, Omsk, Russia}
\date{\today}

\begin{abstract}
Static flexoelectric effect in a finite sample of a solid is addressed in terms of phenomenological theory for
the case of a thin plate subjected to bending.
It has been shown that despite an explicit asymmetry inherent to the bulk constitutive electromechanical equations
which take into account the flexoelectric coupling,  the electromechanical response
for a finite sample is "symmetric".
"Symmetric" means that if a sensor and an actuator are made of a flexoelectric element, performance
of such devices can be characterized by the same effective piezoelectric coefficient.
This behavior is consistent with the thermodynamic arguments offered earlier, being in conflict with the current point of view on the matter in literature.
This result was obtained using standard mechanical boundary conditions valid for the case where the polarization vanishes
at the surface.
It was shown that, for the case where there is the polarization is nonzero at the surface, the aforementioned
symmetry of electromechanical response may be violated if standard mechanical boundary conditions are used, leading to a conflict with the thermodynamic arguments.
It was argued that this conflict may be resolved when using modified mechanical boundary conditions.
It was also shown that the contribution of surface piezoelectricity to the flexoelectric response of a finite
sample is expected to be comparable to that of the static bulk contribution (including the material with high
values of the dielectric constant) and to scale as the bulk value of the dielectric constant (similar to the bulk contribution).
This finding implies that if the experimentally measured flexoelectric coefficient scales as the dielectric
constant of the material, this \emph{does not imply} that  the measured flexoelectric response is controlled by the static
bulk contribution to the flexoelectric effect.

\end{abstract}


\maketitle

\section{Introduction}
The flexoelectric effect consists of a linear response of the dielectric polarization to a strain gradient.
This is a high-order electromechanical which is expected, in general, to be rather weak.
However, some of its features make this effect to be of interest from both fundamental and applied points of view.
This has stimulated recent intensive experimental \cite{Ma,Cross,Zubko}and theoretical
\cite{Sharma,Maranganti,Eliseev}activity in the field.

On the fundamental side, it is of interest that this effect cannot be considered just as a non-local generalization of the piezoelectric effect.
In contrast to the later, the flexoelectric effect (response) is controlled by 4 mechanisms of different
physical nature, the contributions of which can be comparable\cite{Tag1985}.

On the practical side, of first importance is that this effect, in contrast to the piezoelectric effect, is
allowed in centro-symmetric material.
It is believed that it is the flexoelectric effect that is responsible for the generation of an electric field
in acoustic shock waves propagating in centro-symmetric solids\cite{Harris}.
It was recently shown that this effect plays also an essential role in  electromechanical properties of materials
with a moderate level of electronic and ionic conductivity\cite{Moroz2011}.
However, the most applied interest is focused on the "piezoelectric metamaterial" -- composites made of
non-piezoelectric components, which exhibit effective piezoelectric response generated due to the flexoelectric effect.
The work in this direction was initiated by pioneering experimental studies by Professor Cross with coworkers
\cite{Ma,Cross} and was later also supported  by theory \cite{Sharma}.
Presently, $\text{(Ba,Sr)TiO}_3$-based composites have been shown to yield effective piezoelectric
coefficients comparable to those of commercial piezoelectric ceramics \cite{Chu}.
It was argued, based on the constitutive equation for the flexoelectric response, that a mechanical sensor made
of such metamaterials should exhibit a very unusual property.
Specifically, in contrast to piezoelectric based devices, it will not behave as an actuator\cite{Chu,Cross}.
There are several reasons to question such statement.
First, already in the 60's of the past century,  the group of Professor Bursian reported experimental data on
$\text{BaTiO}_3$ crystals\cite{Bursian1968} and gave arguments based on  equilibrium thermodynamics\cite{Bursian1974}, which contradict this statement.
Second, the existence of a linear sensor-not-actuator may come into conflict with the general principles of thermodynamics.

The goal of this paper is to address theoretically this conflict situation to demonstrate that despite an explicit
asymmetry of the constitutive equations for the bulk flexoelectric effect, when this effect is characterized in
a realistic finite sample the apparent asymmetry of the  electromechanical response will vanish.
In particular, this  implies that the aforementioned piezoelectric metamaterial should exhibit the identical
piezoelectric constants when characterized in "direct" and "converse" regimes.
We will demonstrate this for two leading contributions to the static flexoelectric response: the contribution
of static bulk ferroelectricity and that of surface piezoelectricity.

\section{Static bulk flexoelectricity}
\label{bulk}

In the present paper, being interested in the static or quasi-static situation, we are facing three contributions
to the flexoelectric response, which are associated  with  (i) static bulk ferroelectricity,  (ii) surface
piezoelectricity, and (iii) surface flexoelectricity \cite{Tag1985}.
In this section we will discuss the problem of the relation between the direct and converse effects for
contribution (i), reserving the next section for contribution (ii).
In view of vanishing practical importance of the surface flexoelectric effect (it is expected not to be
enhanced on high-dielectric-constant materials which are the only ones suitable for applications) we will not be
treating it in this paper.

The static bulk flexoelectric effect is customarily described by the following free energy (density) expansion
(see e.g.\cite{Eliseev}):
\begin{equation}\label{F}
F = \frac{\chi^{-1}_{ij}}{2} P_i P_j -
\frac{\\f_{ijkl}}{2} \left(P_k\frac{\partial u_{ij}}{\partial x_l} -
u_{ij}\frac{\partial P_k}{\partial x_l}\right) + \frac{c_{ijkl}}{2} u_{ij}u_{kl}
\end{equation}
where $P_i$ and $u_{ij}$ are the polarization vector and strain tensor, respectively, and where the Einstein summation
convention is adopted.
We will consider this thermodynamic potential as having the differential $dF=E_idP_i + \sigma_{ij}du_{ij}$.
Then calculating the electric field, $E_i$, and stress tensor, $\sigma_{ij}$, as variational derivatives
of the free energy of the sample, given by the integral of $F$ over its volume, one arrives at the following
constitutive equations:
\begin{equation}\label{E}
E_k=\chi^{-1}_{kj}P_j - f_{ijkl}\frac{\partial u_{ij}}{\partial x_l}
\end{equation}
and
\begin{equation}\label{sigma}
\sigma_{ij}= f_{ijkl}\frac{\partial P_k}{\partial x_l}+ c_{ijkl}u_{kl}.
\end{equation}
The first equation describes a linear polarization  response to strain gradient (direct flexoelectric effect).
The second one describes the converse flexoelectric effect, implying that to get a "mechanical yield", spatial inhomogeneity of the polarization is needed.
From this, one might infer (as customarily done in relevant papers) that the application of a homogeneous
electric field to a sample will not lead to its deformation.
Even being nearly evident, in reality, the last statement is not correct.

\begin{figure}[h]
\includegraphics[width=8cm, keepaspectratio]{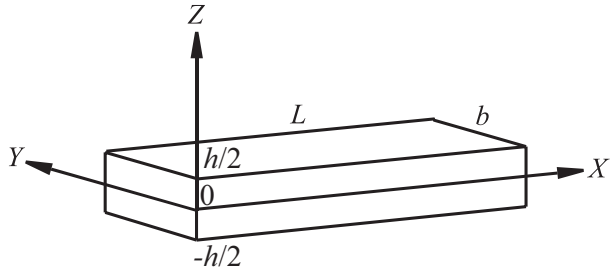}
\caption{Plate of the material exposed to bending and the reference frame used in calculations.}
\label{fig1}
\end{figure}

Let us show this for the flexural mode.
Consider, a (001) plate of a cubic material of thickness $h$ in the reference frame specified in Fig.\ref{fig1},
with the $X$ and $Y$ dimensions being $L$ and $b$ respectively.
To make the analysis transparent, we allow only a cylindrical bending of the plate about $OX_2$ axis.
To simplify the discussion further, we set, for the moment, $c_{1122}=c_{1133}=0$.
In such simplified model, the plate bending is associated with  ${\partial u_{11}}/{\partial x_3}\neq0$,
whereas $u_{22}=u_{33}=0$ so that Eq.\eqref{E} suggests the appearance of $P_3$ component of the polarization
controlled by $f_{1133}$ component of the flexoelectric tensor.
To address the reversibility of this effect, one should check if the application of an electric field normal
to the plate will cause its bending.
A straightforward way do this is to derive the equation of balance of the bending  moment for the
plate\cite{Tim} subjected to a homogeneous electric field $E$ normal to its suface by integrating Eq.\eqref{sigma}
across a $YZ$ cross-section of the sample:
\begin{equation}\label{Mom}
b\int_{-h/2}^{h/2}\sigma_{11}zdz =
bf_{1133}\int_{-h/2}^{h/2}\frac{\partial P_3}{\partial z}zdz+ bc_{1111}\int_{-h/2}^{h/2}u_{11}zdz.
\end{equation}
At mechanical equilibrium,  the lhs term must be equal to the minus the  component of the mechanical moment
of the external forces applied to the lefthand (with respect to the  cross-section of the integration) part
of the plate,$-M_2$.
Without the first rhs term, this equation describes that bending of the sample caused by this moment.
To identify the role of this term, we first evaluate it  using integration by parts:
\begin{equation}\label{Int}
\int_{-h/2}^{h/2}\frac{\partial P_3}{\partial z}zdz =
-\int_{-h/2}^{h/2} P_3dz  = -h\langle P_3 \rangle
\end{equation}
where $\langle P_3 \rangle$ is the averaged polarization induced by the field $E_3$ in the bulk of the plate.
In doing so we assume that the polarization changes continuously from its bulk value to zero
on the plate boundary.
If, however, one explicitly considers nonzero  polarization at the sample surface (as was done previously
 in Ref. \cite{Eliseev}), then one should revise the traditional boundary conditions of the elasticity
theory\cite{Yurkov}.
We will return to this issue later in the paper.
Since the spatial scale of the polarization variation at the interface is much smaller than the thickness of the plate,
with a good accuracy $\langle P_3 \rangle \approx P$, where $P$ -- polarization in the bulk.
Thus, the equation for the moment balance can be rewritten as
\begin{equation}\label{Mom1}
-M_2/b+f_{1133}hP =c_{1111}\int_{-h/2}^{h/2}u_{11}zdz.
\end{equation}

It is clear from this equation that the application of a homogeneous electric field to the plate is equivalent
to that of an external bending moment.
Thus, we conclude  that a finite, mechanically free ($M_2=0$) sample, placed in a homogeneous electric field, will be bent.
This conclusion is closely related to that drawn by Eliseev et al\cite{Eliseev}.
These authors have shown that a ferroelectric plate with the out-of-plane orientation of the spontaneous
polarization should exhibit  spontaneous bending due to the flexoelectric coupling.
It was found that this effect is controlled by a factor $\int_{-h/2}^{h/2}({\partial P_3}/{\partial z})zdz$ which
was calculated using a numerical solution for the polarization profiles $P(z)$ in the sample.
Here it is also worth mentioning that the bending effect addressed, though being proportional to a component of
the \emph{bulk} flexoelectric coefficient and the \emph{bulk} value of the induced polarization, is actually
controlled by forces applied to the surface of the plate.

It is instructive to illustrate quantitatively the "symmetry" of the direct and converse flexoelectric effects
in a finite sample for a situation which is readily mathematically trackable.
We will consider the case of a $(001)$ plate of a cubic  material in  symmetrical flexural mode
with the polarization $P$  normal to the plate and homogenous in its bulk.
In the case of symmetric bending, the curvature of the plate in all crossections normal to it, $G$, is the same.
Enjoying the results of the theory of thin plates \cite{Landau} we can express the components of the strain
tensor in terms of this curvature:
\begin{equation}\label{strain}
u_{11}=u_{22}=zG;~~~ u_{33}=-z\frac{c_{12}}{c_{11}}G;~~~ u_{12}=u_{23}=u_{13}=0.
\end{equation}
Integrating the free energy density, Eq.\eqref{F}, with  strain coming from  Eq.\eqref{strain}, over the
plate thickness, one finds the free energy density per unit area of the plate as a function of $P$ and $G$:
\begin{equation}\label{Fi}
\Phi=\frac{\chi^{-1}_{33}}{2}h P^2 + \frac{D_s}{2}G^2 - 2hPG(f_{1133} - \frac{c_{12}}{c_{11}}f_{1111}).
\end{equation}
\begin{equation}\label{Ds}
D_s= \frac{h^3}{6}\frac{c_{11}^2+c_{11}c_{12}-2c_{12}^2}{c_{11}}.
\end{equation}
where $D_s$ is a coefficient controlling  the flexural rigidity of the plate for this kind of bending.
In derivation of Eq.\eqref{Fi}, we have again used Eq.\eqref{Int} (this gave a factor of 2 in the coupling
term from this equation).
Similar expression for the free energy of flexoelectric plate in the cylindrical bending mode was offered
by Bursian and Trunov \cite{Bursian1974}, based on purely symmetry arguments.
Minimizing Eq.\eqref{Fi} with respect to $P$ and $G$ one finds the equation for the direct and converse effects
for the plate in symmetric flexural mode:
\begin{equation}\label{G}
G=\frac{2h}{D_s}\mu_{\text{pl}}E .
\end{equation}
\begin{equation}\label{P}
D=P= 2\mu_{\text{pl}}G.
\end{equation}
where the electric displacement, $D$, and
\begin{equation}\label{mu}
 \mu_{\text{pl}}=\chi_{33}\frac{c_{11}f_{1133} - c_{12}f_{1111}}{c_{11}}
\end{equation}
can be treated as an effective flexoelectric coefficient of the plate.
The flexural response, given by  Eq.\eqref{G}, is compatible with the results obtained  by Eliseev
et al \cite{Eliseev} for the case of spontaneous bending of thin plates with the blocking boundary
condition for the polarization.

Obviously, elements of such plate will work as both actuators and sensors.
If round pieces of the plate with central loading and symmetric free-edge side support are used as elements
of a piezoelectric metamaterial,  electromechanical properties of the latter will be characterized by a
single effective piezoelectric coefficient $d_{33}$.
Using the relation between the cross-section curvature, $G$, and the maximal deflection, $\xi_{\text{max}}$,
for symmetric bending of a circular plate:
\begin{equation}\label{xi}
 \xi_{\text{max}}= \frac{GR^2}{2}.
\end{equation}
(where $R$ is the radius of the plate) and using Eq.\eqref{G} one readily finds
\begin{equation}\label{d33}
 d_{33}= \frac{\mu_{\text{pl}}R^2}{D_s}.
\end{equation}

\section{Contribution of surface piezoelectricity}
\label{surface}

As was recognized at the first thorough treatment of the flexoelectric response \cite{Tag1985}, the polarization
response to a strain gradient in a finite sample, generally speaking, may not be fully controlled by the
contribution of  the bulk static flexoelectricity, even in materials with high values of the dielectric
constant (high-$K$ materials).
The competing effect that is due to surface piezoelectricity, was not, however, properly addressed theoretically.
Thus, it is not clear if it can in fact compete in  high-$K$ materials with the static bulk flexoelectricity.
In  high-$K$ materials, the contribution of the static bulk flexoelectricity is enhanced, since it scales as
the dielectric susceptibility (cf. Eq.\eqref{P} and Eq.\eqref{mu}).
At the same time, the effect associated with  surface piezoelectricity originates from the presence of
the interface adjacent layers where the piezoelectricity is induced by the inversion-symmetry-breaking effect
of the interface.
Since the sign of the effective piezoelectric coefficients of the layers on the opposite sides of a plate
should be opposite (as controlled by the orientation of the surface normal), bending of the plate should result
in dipole moments in these layers, the sign of which are the same.
The dipole moment in a layer is  proportional to the strain in it, which, in turn, is proportional to the product
of the strain gradient and the plate thickness.
Having calculated the resulting change of the average polarization of the whole system,  this will give rise to
a net polarization proportional to the strain gradient.
From this reasoning it is not obvious that such response will be enhanced once that the dielectric constant of
the bulk of the material is high.
However, such reasoning does not provide a proper vision of the whole effect.
In what follows, we will show that such enhancement does take place and  the considered bulk and surface contributions
to the flexoelectric response can be readily comparable in  high-$K$ materials.
We will also address the problem of the relation between the direct and converse effect for the mechanism
related to the surface piezoelectricity.

\begin{figure}[h]
\includegraphics[width=8cm, keepaspectratio]{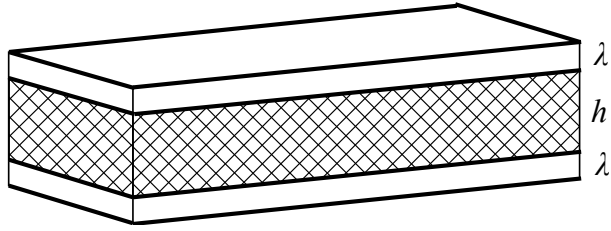}
\caption{Model for the contribution of  surface piezoelectricity to the flexoelectric response of a non-piezoelectric material.
The surface layers of thickness $\lambda$ model the surface adjacent (atomically thin) layers of the material where the piezoelectricity is induced by the symmetry breaking impact of the surface.}
\label{fig2}
\end{figure}

To be specific we will address these problems for the case of symmetric bending of a thin $(001)$ plate of a cubic material.
We will model the effect by considering a system consisting of a plate of an "ideally homogenous"  material
(i.e. its material parameters are the same throughout the plate) with the bulk flexoelectric effect being neglected
and two  thin surface piezoelectric layers (Fig.\ref{fig2}).
The thickness of each layer, $\lambda$, is much smaller that that of the  plate, $h$.
The top layer is characterized by  piezoelectric moduli $h_{333}$ and $h_{311}=h_{322}$, whereas for the bottom
layer these moduli have the same the absolute value but are of opposite sign.
We also ascribe to these layers an out-of-plane component of the dielectric constant, $\varepsilon_\lambda$.

Let us find the extra free energy associated with the top piezoelectric layer when the plate is symmetrically bent with
a cross-sectional curvature, $G$, and when out-of-plain polarization in the layer equals $P_{\lambda}$.
We assume no difference in the elastic properties of the piezoelectric layers and the plate.

We start with the free energy density in the layer defined as
\begin{equation}\label{FL}
F_{\lambda} = \frac{\alpha}{2} P_{\lambda}^2 -
P_{\lambda}[h_{333}u_{33}+h_{311}(u_{11}+u_{22})]+
\frac{c_{11}}{2}(u_{11}^2+u_{22}^2+u_{33}^2)+ c_{12}(u_{11}u_{33}+u_{22}u_{11}+u_{22}u_{33})
\end{equation}
where $c_{11}=c_{1111}$, $c_{12}=c_{1122}$, and $\alpha$ is the inverse dielectric susceptibility of the layer
if it were fully mechanically clamped.
Because the plate is thin we set in the layer
\begin{equation}\label{strainL}
u_{11}=u_{22}=hG/2.
\end{equation}
As for $u_{33}$, we find it from the condition that the surface of the film is mechanically free,
$\partial F_{\lambda}/\partial u_{33}=0$:
\begin{equation}\label{u33}
u_{33} = -\frac{c_{12}}{c_{11}}hG + \frac{h_{333}}{c_{11}}P_{\lambda}.
\end{equation}
Inserting Eqs.\eqref{strainL} and \eqref{u33}) into Eq.\eqref{FL} and  multiplying the result with $\lambda$,
one finds the free energy of top piezoelectric layer.
The energy of the bottom piezoelectric layer is the same since this layer differs from the top layer by the sign
of piezo-moduli and by that of the strain; in the expression for energy these signs cancel each other out.
Thus, for the contribution of the two piezoelectric layers to the free energy of the system, we find:
\begin{equation}\label{FIL}
\Phi_{\lambda} = 2 \lambda \left[\frac{\chi_{\lambda}^{-1}}{2} P_{\lambda}^2 -
(h_{311}-\frac{c_{12}}{c_{11}}h_{333})hGP_{\lambda}\right]+\Phi_1
\end{equation}
where $\chi_{\lambda} = (\alpha - h_{333}^2/c_{11})^{-1}$is the true (under the mixed mechanical conditions)
dielectric susceptibility of the piezoelectric layers and $\Phi_1$ is their mechanical bending energy.

To describe the direct flexoelectric response we use the equation of state for the polarization in the
piezoelectric layer, $\partial \Phi_{\lambda}/\partial P_{\lambda}=E_{\lambda}$ ($E_{\lambda}$ is
the electric filed in the layer), the condition of continuity of the electric displacement, $D$,
in the layer, and the short-circuit condition.
This leads to the following set of equations:
\begin{equation}\label{dir1}
P_{\lambda} = \chi_{\lambda}E_{\lambda}+ ehG
\end{equation}
\begin{equation}\label{dir2}
D=\varepsilon_f E_f = \varepsilon_0 E_{\lambda}+P_{\lambda}
\end{equation}
\begin{equation}\label{dir3}
2\lambda E_{\lambda} + h E_f = 0
\end{equation}
where $E_f$, $\varepsilon_f$, $\varepsilon_0$ are the electric field in the bulk of the plate, its dielectric
constant, and the dielectric constant of the free space, respectively, and
\begin{equation}\label{dir4}
e= \chi_{\lambda}(h_{311}-\frac{c_{12}}{c_{11}}h_{333}).
\end{equation}
Solving this set of equations we find the relation for the direct flexoelectric response:
\begin{equation}\label{D}
D= 2 \widetilde{e}\lambda G.
\end{equation}
where
\begin{equation}\label{e}
\widetilde{e}= e\frac{h \varepsilon_{f}}{2\lambda \varepsilon_{f} +h \varepsilon_{\lambda}}
\end{equation}
and $\varepsilon_{\lambda}=\varepsilon_{0}+\chi_{\lambda}$.

To describe the converse flexoelectric response we present the elastic energy of the system at fixed
$P_{\lambda}$ in the form
\begin{equation}\label{Fi1}
\Phi= \frac{D_s}{2}G^2 - 2h \lambda e G E_{\lambda}.
\end{equation}
When writing this equation we have neglected the elastic energy of the surface layer, $\Phi_1$, compared
to that of the plate and have simplified \eqref{dir1} down to  $P_{\lambda} \approx \chi_{\lambda}E_{\lambda}$.
This approximation means that we neglect feedback effect of $G$ on $P_{\lambda}$.
This effect will yield some renormalization
of $D_s$, but  practically such a renormalization is negligible indeed.
Minimizing $\Phi$ with respect to $G$ and applying the electrostatic relation used above we  arrive
at the set of equations
\begin{equation}\label{con1}
G = \frac {2h \lambda e}{D_s}E_{\lambda}
\end{equation}
\begin{equation}\label{con2}
\varepsilon_f E_f = \varepsilon_{\lambda}E_{\lambda}
\end{equation}
\begin{equation}\label{con3}
2\lambda E_{\lambda} + h E_f = E(h+2\lambda)
\end{equation}
where $E$ is the applied (average) electric field.
This set leads us to the equation for the converse flexoelectric effect in the system:
\begin{equation}\label{G1}
G=\frac{2h}{D_s}\widetilde{e}\lambda E .
\end{equation}
When writing this relation, only the leading terms to within a small parameter $\lambda/h$ were kept.

The following remarks are to be made concerning the results obtained.
First, the relations obtained for the contribution of the surface piezoelectricity into the flexoelectric
response, Eqs. \eqref{D} and \eqref{G1}, are identical to those obtained for the case of  static bulk
flexoelectricity, Eqs. \eqref{P} and \eqref{G},  to within the replacement
$\mu_{\text{pl}}\Rightarrow \widetilde{e}\lambda$.
Thus, all conclusions about the direct-converse-effect symmetry drawn in the previous section for the static
bulk flexoelectricity hold for the contribution associated with  the surface piezoelectricity.
Second, in a high-K material the latter contribution scales as its dielectric constant (similar to the case
of the static bulk flexoelectricity).
Formally,  this follows from the expression for $\widetilde{e}$, Eq. \eqref{e}.
Taking into account that the thickness of the surface piezoelectric layer is expected to be of the order of
the lattice constant, for realistic values of the plate thickness, $\widetilde{e}$ can be evaluated as
\begin{equation}\label{e1}
\widetilde{e}= e\frac{\varepsilon_{f}}{\varepsilon_{\lambda}}.
\end{equation}
Note that there is no reason to consider $\varepsilon_{\lambda}$ in a high-K material (typically it is
a "regular" of incipient ferroelectric in the paraelectric phase) to be high, since the special interplay of
the atomic forces responsible for the high value of the bulk permittivity will be inevitably destroyed
in the surface layer.
Such enhancement of a surface driven effect by a factor of the bulk permittivity looks surprising.
However, the physical mechanisms behind this effect can be identified.

For the converse effect it is quite transparent.
The bending of the system is controlled by the value of the field in the surface layer
(cf. Eq. \eqref{con1}). Due to its small thickness, this field is enhanced by a factor of
$\varepsilon_{f}/\varepsilon_{\lambda}$ , compared to the applied field.

For the direct effect, the explanation is less straightforward.
This time, the bending creates polarization in the surface layer.
Because of the inhomogeneity of the system, the short-circuiting does not guarantee the absence of the electric
field in it so that the polarization in the surface layer induces a depolarizing field both in itself and
in the bulk of the plate.
It occurs that if the surface layer is thin enough whereas $\varepsilon_{f}/\varepsilon_{\lambda}$ is large,
the polarization response is controlled by the depolarizing field in the bulk of the plate.
This way the polarization response of the system becomes sensitive to the bulk value of the dielectric constant.

Another important conclusion is that, taking into account the aforementioned effect of enhancement,
one expects both contributions to the flexoelectric response discussed to be of the same order of magnitude
even in high-K materials.
These contributions would be comparable if $\lambda \widetilde{e}/\varepsilon_{f}$ were about the typical
value of the components of the flexoelectric tensor $f_{ijkl}$ , $1-10~\text{V}$ (see e.g.\cite{Eliseev}).
For the "atomic values" of the entering parameters ($ \lambda=0.4 ~\text{nm}$ and $e =1~\text{C}/ \text{m}^2$)
and  $\varepsilon_{\lambda}/\varepsilon_{0}=10$,  we evaluate
$\lambda \widetilde{e}/\varepsilon_{f}\simeq 4~\text{V}$ to find that this is, in fact, the case.

\section{Discussion and conclusions}

The analysis presented clearly demonstrates that starting from continuous constitutive electromechanical equations
one can derive the relations for the direct and converse flexoelectric effects  in a finite sample, which exhibit
the symmetry required by thermodynamics.
On the practical level such symmetry implies that a piezoelectric meta-material based on the flexoelectric effect
will exhibit the same effective piezoelectric coefficient in the testing regimes for the direct and converse
piezoelectric effect.
This is in conflict to the belief of those dealing with such meta-materials \cite{Chu,Cross}.

At the same time, it is also clear that the analysis, based on the constitutive electromechanical equations
presented above, is valid only for the situation where the polarization at the plate surface can be treated as
continuously changing from its bulk value to zero.
If it is not the case, formally following this analysis one readily finds that the aforementioned symmetry is violated.
For example, for the case of free boundary conditions for polarization (${\partial P}/{\partial z}=0$
at the boundary), corresponding to nonzero  polarization  at the sample surface, the equation of mechanical
equilibrium, Eq. \eqref{Mom}, implies the absence of the converse effect.
Such conclusion would be fully consistent with that by Eliseev et al \cite{Eliseev} who argued that
the manifestation of the converse flexoelectric effect in a  plate is strongly dependent on the boundary conditions
for the polarization.
At the same time, there is no reason to expect that the free  boundary conditions for the polarization will suppress
the direct flexoelectric effect.
Thus, if we followed the  calculating scheme employed by Eliseev et al \cite{Eliseev} (and used
in Sect.\ref{bulk}) we would find, for the  free polarization boundary conditions, the absence of the converse
flexoelectric effect in the presence of the direct effect.
This would make an apparent contradiction between the results obtained from the continuous constitutive equations
and those obtained from thermodynamics.

We suggest the following resolution to this contradiction.
The point is that incorporating the flexoelectric coupling into the free energy density of a material leads to
a modification of the boundary conditions for the bulk constitutive equations.
Eliseev et al \cite{Eliseev} have derived modified boundary conditions for the polarization,
however these authors have postulated that the classical mechanical boundary conditions are not affected
by such incorporation.
However, as was recently shown by one of the authors\cite{Yurkov}, generally, the mechanical boundary
conditions should be modified as well.
It has been shown that such boundary conditions reduce down to the classical mechanical boundary conditions
when the polarization vanishes at the surface.
This justifies the calculations based on the classical mechanical boundary conditions, which we have presented
in Sect.\ref{bulk}.
For the  general case, the problem of the converse flexoelectric effects should be revisited with the
correct mechanical boundary conditions which contain the surface value of the polarization.
We expect that  such treatment will yield the results consistent with the symmetry between direct and converse
flexoelectric effects in a finite sample dictated by thermodynamic arguments.

Another important conclusion follows from the results obtained in Sect.\ref{surface}.
There, it was shown that the contribution of the surface piezoelectricity to the flexoelectric response is
expected to be comparable to that of the static bulk contribution (including the material with high values
of the dielectric constant) and to scale with the bulk value of the dielectric constant
(similar to the bulk contribution).
The latter statement actually implies identical (or at least similar) temperature dependences of these contributions.
Note that in earlier publications \cite{Tag1985} it was hypothesized that these depedences are expected to be different.
Based on this hypothesis, the fact that the experimentally measured flexoelectric coefficient scales
as the dielectric constant of the material was customarily taken as an indication that the measured flexoelectric
response is controlled by the static bulk contribution to the flexoelectric effect.
The results from Sect.\ref{surface} essentially change the situation.
Now one can state that the fact that the experimentally measured flexoelectric coefficient scales as
the dielectric constant of the material \emph{does not imply} that  the measured flexoelectric response
is controlled by the static bulk contribution to the flexoelectric effect.

\section{Acknowledgements}\label{Acknowledgments}
This project was supported by Swiss National Science Foundation.
Andrey Zakurdaev is acknowledged for reading the manuscript.

\end{document}